# A Theory for Efficient Coding of a Dynamic Trajectory Predicts non-Uniform Allocation of Grid Cells to Modules in the Entorhinal Cortex

## Authors

Noga Weiss Mosheiff[1], Haggai Agmon[2], Avraham Moriel[1], and Yoram Burak[1,2]

[1] Racah Institute of Physics, Hebrew University, Edmond J. Safra Campus, Jerusalem, Israel

[2] Edmond and Lily Safra Center for Brain Sciences, Hebrew University, Edmond J. Safra Campus, Jerusalem, Israel

## Correspondence

yoram.burak@elsc.huji.ac.il

September 20, 2015

## SUMMARY

Grid cells in the entorhinal cortex encode the position of an animal in its environment using spatially periodic tuning curves of varying periodicity. Recent experiments established that these cells are functionally organized in discrete modules with uniform grid spacing. Here we develop a theory for efficient coding of position, which takes into account the temporal statistics of the animal's motion. The theory predicts a sharp decrease of module population sizes with grid spacing, in agreement with the trends seen in the experimental data. We identify a simple scheme for readout of the grid cell code by neural circuitry, that can match in accuracy the optimal Bayesian decoder of the spikes. This readout scheme requires persistence over varying timescales, ranging from ~1ms to ~1s, depending on the grid cell module. Our results suggest that the brain employs an efficient representation of position which takes advantage of the spatiotemporal statistics of the encoded variable, in similarity to the principles that govern early sensory coding.

**INTRODUCTION**

A central goal of systems neuroscience is to unravel the principles of encoding in the brain. In primary sensory areas, it has been conjectured that the neural circuitry implements coding schemes that maximize information about sensory inputs, while constraining neural resources such as the number of cells or the rate of spikes. This hypothesis (Barlow, 1961) has been particularly successful in explaining neural responses in early visual and auditory areas (Atick and Redlich, 1992; Bell and Sejnowski, 1997; Fairhall et al., 2001; Laughlin, 1981; Olshausen and Field, 1997; Smith and Lewicki, 2006). More recently, it was proposed that grid cells in the entorhinal cortex (Hafting et al., 2005) implement an efficient code for an internally computed quantity, the position of an animal in its environment (Fiete et al., 2008; Mathis et al., 2012a; 2012b; Sreenivasan and Fiete, 2011; Wei et al., 2015). According to this proposal, the neural code for position possesses a dynamic range (defined as the ratio between the representable range and the resolution) that depends exponentially on the number of encoding neurons (Burak, 2014). Thus, the dynamic range of the grid cell code vastly exceeds that of unimodal coding schemes, such as the encoding of position in the hippocampus (Hartley et al., 2014), or the encoding in head direction cells (Taube, 2007).

Most theoretical treatments of grid cell coding assumed a uniform distribution of neurons across grid cell modules (Fiete et al., 2008; Mathis et al., 2012b), or deduced that such a distribution is expected based on an optimization principle (Mathis et al., 2012a; Wei et al., 2015) (here, a module is defined as a group of grid cells that share the same grid spacing and orientation). However, in a systematic characterization of grid cell parameters (Stensola et al., 2012), many more cells were found in modules with small spacing, compared to modules with larger spacing (Fig. 1A). The study by Stensola et al (Stensola et al., 2012) did not attempt to quantify precisely the population sizes in different modules, and the reported numbers were likely influenced by experimental biases, but the trend observed in this study is so pronounced, that it is highly suggestive of a non-uniform distribution of grid cells across modules. This observation raises two questions:  First, is a non-uniform distribution of grid cells across modules compatible with the efficient coding hypothesis? Second, if the number of cells in modules with large spacing is indeed very small as hinted by the existing experimental evidence, can these cells support at all a precise neural representation of position?

Here we propose that the entorhinal cortex is adapted to represent a *dynamic* quantity, the animal's trajectory in its environment, while taking into account the temporal statistics of this variable in addition to the requirements on spatial range and resolution. This hypothesis leads to the prediction that grid cell population sizes should decrease sharply as a function of grid spacing, whereas grid spacings should follow approximately a geometric series as predicted previously (Mathis et al., 2012b; 2012a; Wei et al., 2015) – in agreement with experimental evidence (Stensola et al., 2012). We show, in addition, that fairly simple neural circuitry can reliably read out a neural code with these properties, while taking into account the temporal statistics of the animal's location. This reasoning leads to an interesting prediction on the processing of spikes downstream of the entorhinal cortex: the characteristic integration time of spikes in neurons that implement a readout of position is expected to increase sharply as a function of the grid spacing of the presynaptic grid cell.

## RESULTS

### Encoding of a dynamic location

First, we briefly review the theoretical considerations relevant to the representation of a static variable. Imagine an ideal observer that attempts to read out position from the spikes generated by all the neurons in one module with grid spacing $\lambda$, over a time interval $\Delta T$. If the rate of spikes is sufficiently large, the posterior distribution over position is approximately given by a periodic array of Gaussians (Fig. 1B). The spatial periodicity of this distribution is a consequence of the single neuron tuning curves, which all share the same periodic structure. If the individual receptive fields are isotropic and compact, the Gaussians are isotropic as well (Mathis et al., 2015). Their characteristic width reflects the precision at which position can be read out locally within a unit cell of the periodic lattice. For independent Poisson spikes,

(1)

$$\Delta^2 = \frac{2}{J \Delta T}$$

where $\Delta^2$ is the mean variance of the periodic Gaussians, the factor 2 comes from the two dimensions, and $J$, the Fisher information rate (in each direction in space, see Supplemental Information) is independent of $\Delta T$ for Poisson neurons, and can be written as:

(2)

$$J = \alpha \frac{n}{\lambda^2}.$$

Here $n$ is the number of neurons in the module, and the proportionality constant $\alpha$ depends on the detailed shape of the firing fields (see Supplemental Information for a derivation of $\alpha$ for Gaussian receptive fields). We assumed that neurons within the module cover densely and uniformly all possible phases of the periodic tuning curve, and that the Cramér-Rao bound is saturated. The dependence of $J$ on $\lambda$ can be deduced based on dimensional analysis, relying on the observations that the maximal firing rate, $r_{\max}$, is approximately constant in different modules, and that firing fields of grid cells scale in proportion to the grid spacing (Hafting et al., 2005) (therefore, $\lambda$ is the only spatial length scale characterising the response in each module). Note that the precision of readout depends on the choice of the observation interval, and that the MSE is inversely proportional to the number of neurons (Fig. 1D).

Based on Eqs. (1) and (2), a uniform allocation of neurons to modules implies that the ratio $\Delta/\lambda$, the precision of readout relative to the grid spacing, is the same for all modules. Intuitively, this is a plausible requirement, and indeed this relation was postulated (or derived) in previous works: for example, consider a nested coding scheme (Mathis et al., 2012b; 2012a; Wei et al., 2015), in which the grid spacings follow a geometric series. Let us denote by $\lambda_i$ the grid spacings, ordered sequentially ($\lambda_1 > \lambda_2 > \cdots$), and by $\Delta_i$ the corresponding precision of readout from each module. Uniformity of $\Delta_i/\lambda_i$ across modules implies also that the ratio $\Delta_i/\lambda_{i+1}$ is uniform across modules. A sufficiently small value of this ratio ensures that readout from each module is accurate enough to avoid ambiguities arising from the periodicity of response in the successive module with smaller spacing. Thus, by choosing a fixed (and sufficiently small) ratio $\Delta_i/\lambda_{i+1}$, it is possible to ensure that ambiguities do not arise in the readout of the code at any scale.

To see why the dynamic aspect of the trajectory is consequential, let us suppose that the animal's trajectory follows the statistics of a simple random walk. We imagine that each neuron fires as an inhomogeneous Poisson process with a rate determined by the tuning curve of the neuron, evaluated at the instantaneous position of the animal. Consider an ideal observer, attempting to estimate the animal's position at time $t$, based on the spikes from all neurons in a single module, emitted up to that time (Fig. 1C). In the

Supplemental Information we show that the local mean square error (MSE) of such an optimal estimator is given by

(3)

$$\Delta^2 = 2\sqrt{\frac{2D}{J}},$$

instead of Eq. (1) (see also Fig. 1E). Thus the MSE is proportional to $n^{-1/2}$, instead of the $n^{-1}$ dependence of the static case (compare Figs. 1 D and E). This difference in scaling with $n$ may seem minor, but using Eqs. (2) and (3) we find that in order to achieve a fixed relative precision $(\Delta_i/\lambda_i)$ for all modules, it is now necessary to have

(4)

$$n \sim \frac{1}{\lambda^2}.$$

Thus, far fewer neurons are required in modules with large spacing, compared to modules with small spacing. This result can be easily explained in qualitative terms: the relative position of the animal, in relation to the period of the grid response, varies more slowly in the modules with large $\lambda$ compared to modules with small $\lambda$. Thus, an ideal decoder can rely on spikes emitted within a longer period of time in order to estimate the relative position within a unit cell of the periodic grid. The validity of this interpretation is further demonstrated below (*Biological implications for dynamic readout*).

**Optimal module population sizes**

To see how these principles impact a more detailed theory for the allocation of grid cells to modules, we consider a nested code (Mathis et al., 2012a; Wei et al., 2015; Weiss et al., 2012), in which position can be read out sequentially starting from the module with the largest spacing, progressing sequentially to modules with smaller grid spacings. We follow a similar line of argumentation as in (Wei et al., 2015). Our goal is to minimize $\Delta_m$, the resolution of readout from the smallest module, while constraining the largest grid spacing $\lambda_1$ and the number of neurons $N$ (equivalently, it is possible to minimize the number of neurons while constraining the readout resolution). Additionally, we require that ambiguities about position do not arise at any one of the refinement steps. Therefore, we impose a relation between the readout resolution and grid spacing,

(5)

$$\Delta_i = \beta\lambda_{i+1}.$$

Here, $\beta$ should be sufficiently small such that the range of likely positions, inferred from module $i$, does not contain multiple periods of the response from module $i + 1$. Below, the value of $\beta$ is set as described in the Experimental Procedures and Fig. S3. Crucially, we use Eq. (3) for the resolution of readout at each step, since we hypothesize that grid cells encode a dynamic position with random walk statistics. Additional details of the optimization are described in the Experimental Procedures and the Supplemental Information. The requirement of unambiguous reconstruction, combined with Eq. (3), leads to several salient results.

First, we find that in the optimized code, the module population sizes precisely follow a geometric progression

(6)

$$n_i \sim 2^i \, ,$$

where $n_i$ is the number of neurons in module $i$, Fig. 2A. Second, we find that the ratios between subsequent grid spacings are approximately constant in the modules with small spacing. The optimal ratio approaches a limit given by $\sqrt{2} \simeq 1.41$ for the smallest modules (Supplemental Information and Fig. 2C). This prediction is in close agreement with the ratio of grid spacings in subsequent modules, measured in (Stensola et al., 2012) and averaged across animals, approximately 1.42. Note that the ratios were measured only for the first few modules with lowest grid spacings, hence the theory is in very good agreement with the existing measurements.

The properties listed above are independent of the total number of neurons, the shape of the tuning curve, and the parameter $\beta$. Moreover, these properties remain intact even if we relax the assumption of an optimal estimator, but assume the relation $\Delta^2 \sim J^{-1/2}$ , as in Eq. (3).   Other, more detailed aspects of the results do depend on parameters. In Fig. 2 we assumed that the total number of grid cells is either $10^4$ (blue) or $10^5$ (red), leading to differences in the ratios between subsequent modules – but not in the ratios obtained for the smallest modules. Most importantly for our discussion on the allocation of grid cells to modules, the module population sizes are given precisely by Eqs. (6) and (8), irrespective of parameters. In particular, about half of the neurons are allocated to the module with the smallest grid spacing (Fig. 2A).

It may seem surprising that accurate readout is possible at all with only a handful of neurons in the modules with the largest spacing. Figure 3A shows the estimate of position obtained by an optimal Bayesian decoder (see Supplemental Information), in response to simulated spikes from ten modules with the above parameters (altogether

$10^4$ grid cells, and ten neurons in the module with the largest spacing). The root mean square error (RMSE) of this estimator is $1.276 \pm 0.004$ cm. It is instructive to compare this result with the performance under two other allocations of grid cells to modules: if the neurons are allocated with equal proportion to all modules, the RMSE is increased by a factor of about 1.5 (Fig. 3B). If the allocation of neurons to modules is reversed, such that most neurons participate in the modules with larger grid spacing, the RMSE becomes larger by a factor of about 3.4 (Fig. 3C).

In summary, the hypothesis that grid cells are adapted to efficiently encode a dynamic position predicts a sharp decrease in the number of grid cells allocated to modules with high grid spacing, compared to modules with smaller spacing, while remaining compatible with previous theories, which predicted a geometric progression in the grid spacings.

### Biological implications for dynamic readout

The analysis of grid cell activity from the perspective of an ideal observer is relevant for coding in the brain only if neural circuitry can implement an efficient decoding scheme of the grid cell code, while taking into account the statistics of the animal's motion. The direct computations involved in a precisely optimal decoder [Eq. (10)] are elaborate, but we show next that fairly undemanding processing of the spikes can produce near optimal readout of position from each module.

We analyze a simple readout scheme in which spikes emitted by grid cells are interpreted as if the position of the animal is static. For a truly static position, all the spikes emitted in the past are informative about the current position. Here, however, we consider an estimate of position which is constructed based only on spikes from the recent history, weighted by an exponential kernel with time constant $\tau$ (Fig. 4A). An estimator that treats the position of the animal as if it is static, has a simple structure: the log likelihood to be at position $\vec{x}$ can be expressed as a linear function of the spike counts (Supplemental Information). A single non-linearity is then sufficient to select the position $\vec{x}$ at which the log likelihood is maximal.

Since the trajectory of the animal is in fact dynamic, the above estimator is, in general, suboptimal. Its best performance is obtained by choosing $\tau$ as follows (Supplemental Information),

(7)

$$\tau = \frac{1}{\sqrt{2DJ}} = \frac{\lambda}{\sqrt{2D\alpha n}}.$$

This choice balances two contributions to the error of the estimator, with opposing dependencies on $\tau$: first, the ambiguity in the decoding of position due to the stochasticity of spikes, which becomes large when $\tau$ is small (and few spikes contribute to the estimate). The second contribution to the error is due to the animal's motion. This contribution increases with $\tau$, since the simple decoder ignores the animal's motion altogether. In the Supplemental Information we show that despite its simplicity, the above estimator achieves the same performance as the optimal Bayesian decoder, Eq. (3), when the readout time $\tau$ is chosen according to Eq. (7).

According to Eqs. (4) and (7), the time scale $\tau$ should decrease in sequential modules by a factor of 2 for the modules with smaller grid spacings, where Eq. (4) is approximately valid. Characteristic values of $\tau$ are shown in Fig. 4B, where the parameters are the same as in Fig. 2. In this example, $\tau$ varies from ∼1 ms to ∼600 ms, depending on the grid spacing.

Based on these results, we consider a simple model for readout of the grid cell code, in which place cells in the hippocampus approximate the log likelihood of position based on incoming spikes from the entorhinal cortex. We model the activity of each place cell as linearly determined from incoming spikes, with an exponential temporal kernel whose characteristic time constant $\tau$ depends on the grid spacing of the presynaptic input cell, Fig. 4C. A single exponential nonlinearity is then sufficient to obtain an approximation of the likelihood. In addition, lateral inhibitory connectivity in the place cell network, not modeled explicitly here, could implement winner-take-all dynamics (Dayan and Abbott, 2001) which would serve to select a unique estimate for the maximum-likelihood position.

With the readout time constants set by Eq. (7), and with appropriately chosen synaptic weights, selecting the cell with the maximal activation yields an estimate for position with a MSE which is close to that of an optimal Bayesian decoder (compare Fig. 4D and Fig. 3).

An interesting prediction follows for the readout of position in the hippocampus (or in other brain areas), based on inputs from grid cells: Spikes in grid cells are expected to influence the activity of a postsynaptic readout cell over an integration time that depends on the grid spacing. The integration time should increase monotonically with grid spacing, as predicted by Eq. (7).

**Optimization for other trajectory statistics**

Within the above readout scheme, it is possible to adjust the module properties in order to optimize the resolution of readout for trajectories that deviate from simple random walk statistics. Let us suppose, for example, that the variance of motion increases quadratically with time, instead of the linear dependence that characterizes a simple random walk. This scenario corresponds to motion at a constant velocity, and at a random direction which is unknown to the decoder of the spikes. It straightforward to evaluate the readout error of the simple estimator in each module, under this type of motion (see SI), and to find the value of $\tau$ that minimises its MSE, Eq. (17).

We thus repeat our optimization scheme for the number of cells in each module and the grid spacings, while using Eq. (14) instead of the expression for random walk statistics, Eq. (3). The predicted ratio in the number of cells between successive modules is 1.5 instead of 2 (Fig. 5A), the asymptotic ratio between the spacings in successive modules is 1.5 instead of $\sqrt{2} \simeq 1.4$ (Fig 5C), and the range of optimal readout times is somewhat narrower than obtained under the assumption of random walk statistics (Fig. 5D). Nevertheless, the qualitative conclusions are very similar under the two scenarios: module population sizes decrease sharply with grid spacing, and the ratios of successive grid spacings are approximately constant (and similar in the two scenarios) for the modules with small spacing. Thus, the qualitative conclusions are valid for a wide range of statistics that may characterize the motion.

**DISCUSSION**

In summary, we propose that the representation of position in the entorhinal cortex takes advantage of the continuous temporal statistics of motion, in order to efficiently encode the animal's position. This is possible due to the multiscale structure of the code: in modules with larger grid spacing, the encoded variable varies less rapidly than in modules with smaller spacing. Spikes from these modules remain informative about the current position over a longer time scale, allowing for an efficient encoder to allocate a smaller number of cells to these modules. The relevance of this argumentation [and related reasoning (Mathis et al., 2012a; Wei et al., 2015)] to the encoding of position in the entorhinal cortex, is suggestive of an intriguing possibility, that the principle of efficient coding (Barlow, 1961) extends in the brain well beyond the realm of early sensory processing.

From the theoretical perspective, it is interesting to consider the dynamical range of the representation: the ratio between the represented range and a measure of the resolution, such as the MSE (Burak, 2014). Optimizing this quantity is a difficult theoretical problem, even when assuming that the encoded variable is static (Mathis et al., 2012a; Wei et al., 2015). Our goal here was not to fully solve this problem, but to explore the salient consequences, arising from a hypothesis that the code is adapted to the dynamics of the animal's trajectory. We focused our analysis on nested codes, and assumed that the range of representation of the grid cell code matches the largest grid spacing. However, we expect that the principles revealed here for the neural representation of a dynamic trajectory apply also if the range of positions encoded by grid cells is much larger (Fiete et al., 2008).

The most important prediction arising from our hypothesis, is the highly non-uniform distribution of grid cells across modules. Previous experiments (Stensola et al., 2012) strongly hint that this is indeed a prominent feature in the organization of the entorhinal cortex, but additional experiments are necessary in order to establish this conclusion more firmly, and to obtain quantitative estimates of the distribution.

Another intriguing prediction arises from the identification of a relatively simple decoding scheme that takes into account the dynamic aspect of motion: action potentials of grid cells are expected to affect the activity of postsynaptic readout cells over varying time scales, which increase with the grid spacing. Predicted time scales span about three orders of magnitude, from ∼1ms to ∼1s, assuming that the largest grid spacing spans a few meters. Integration time scales up to ∼100ms can be implemented in neural circuitry by the dynamics of synaptic integration. Longer time scales of integration in the order of 1s require other mechanisms for persistence (Barak and Tsodyks, 2014): these can potentially rely on recurrent connectivity, on short term synaptic plasticity (Mongillo et al., 2008) or perhaps on intrinsic cellular persistence. It is noteworthy that intrinsic persistence, with characteristic time scales of seconds has been widely observed in the hippocampal formation, and specifically in the entorhinal cortex (Egorov et al., 2002) and the hippocampus (Knauer et al., 2013).

Our model for readout of grid cell activity was deliberately simplified, in order to emphasize the main principles governing the readout of a dynamic variable. Thus, we described the readout as occurring in a single feedforward layer. We speculate that the functional organization along the dorso-ventral axis of the hippocampus may be helpful in implementing different time scales for integration at different spatial scales. Furthermore, several lines of experimental evidence suggest that place cells are driven by environmental sensory inputs that are independent of grid cell activity (Bush et al., 2014; Hales et al., 2014). Nevertheless, there is also compelling evidence that grid cells contribute to the activity of place cells, perhaps most prominently when direct sensory cues are absent, and that the MEC and hippocampus form together a processing loop responsible jointly for spatial representation, computation, and memory (Bush et al., 2014; Hales et al., 2014) [see, also, (Sreenivasan and Fiete, 2011) for a discussion of possible implications from a theoretical perspective].

In short term memory networks, the fidelity of the neural code is consequential for self-maintenance of the persistent state, in addition to its significance for downstream readout (Burak and Fiete, 2012). If the entorhinal cortex autonomously maintains short-term memory of position, as required for idiothetic path integration (Burak and Fiete, 2009; Hafting et al., 2005; McNaughton et al., 2006), we hypothesize that recurrent connectivity within the entorhinal cortex may follow principles similar to those proposed here for readout in the hippocampus.

**EXPERIMENAL PROCUDURES**

**Optimized code**

Details of the optimization are provided in the Supplemental Information. The optimal number of neurons in a module $n_i$ (Fig.2A), and the ratio $\frac{\lambda_i}{\lambda_{i+1}}$ (Fig.2C) are given by:

(8)

$$n_i = \frac{\left(\frac{1}{2}\right)^{m+1-i}}{1 - \left(\frac{1}{2}\right)^m} N \, ,$$

where $N$ is the total number of grid cells, and

(9)

$$\frac{\lambda_i}{\lambda_{i+1}} = \lambda_1 \left(\frac{1}{2}\right)^i \left[\frac{8D}{\alpha N \beta^4}\left(1 - \left(\frac{1}{2}\right)^m\right)\right]^{-\left(\frac{1}{2}\right)^{i+1}} 2^{\left[-2^{-(i+1)}(m+2)+\frac{1}{2}\right]} .$$

For simplicity, we assume Gaussian receptive fields, for which the factor $\alpha = \frac{4}{\sqrt{3}}\pi r_{\max}$ in Eq. (2) (see SI).

The parameter $\beta$ [Eq. (5)] was chosen as follows. This parameter should be set small enough to ensure that there are no global ambiguities, since the minimization of $\Delta_m$ affects only the local inference error. We applied the minimization procedure with various values of $\beta$ to select $n_i$ and $\lambda_i$, and then evaluated the MSE of the exponential kernel estimator. As expected, very small values of $\beta$ led to large MSE, since $\beta$ sets a limit on the degree of reduction in $\Delta$ from module to module. Large values of $\beta$ also led to large MSE due to errors arising from global ambiguities (Fig. S3). We chose in all simulations $\beta$=0.1, as this value provides an MSE close to minimal (we did not attempt to find the precise optimal value of $\beta$, but note that the choice of this parameter does not affect any of the qualitative results).

**Optimal Bayesian decoder**

The posterior probability distribution used by the optimal Bayesian decoder (Fig. 3), is obtained using the dynamic update rule:

(10)

$$p(\vec{x}, t + dt) = \frac{1}{Z}[\int d\vec{x}' p_D(\vec{x}|\vec{x}')p(\vec{x}', t)]p_{spikes}(\vec{x}, t) ,$$

where $p_D(\vec{x}|\vec{x}')$ is the probability for the random walk to reach $\vec{x}$ at time $t + dt$ from position $\vec{x}'$ at time $t$, and $p_{spikes}(\vec{x}, t)$ represents the likelihood of the spikes observed within the short time interval, given the position $\vec{x}$ (see SI for more details). The optimal Bayesian decoder estimates the location of the animal by:

(11)

$$\hat{x}_{\mathrm{ML}} = \mathrm{argmax}_{\vec{x}} p(\vec{x}, t) .$$

**Exponential kernel decoder**

The temporal exponential kernel decoder estimates the location of the animal as follows:

(12)

$$\hat{x} = \text{argmax}_{\vec{x}} \sum_i \sum_{\mu \in i} \int_0^\infty dt' h_i(t') \xi_\mu(t - t') \log \left[ f_i(\vec{x} - \vec{x}_\mu) \right],$$

where the second sum is over neurons $\mu$ that belong to module $i$, $f_i(x)$ is the shape of the tuning curve, characterising the receptive field of the neurons in the $i$th module, $\vec{x}_\mu$ is the center of the receptive field of neuron $\mu$, $\xi_\mu(t)$ is a series of delta functions that represents the spikes of neuron $\mu$, and $h_i(t)$ is the temporal kernel of module $i$. In our case:

$$h_i(t) = \exp\left(-\frac{t}{\tau_i}\right).$$

The posterior probability distribution illustrated in Fig. 4D is given by

(13)

$$p(\vec{x}, t) = \frac{1}{Z} \exp\left\{ \sum_i \sum_{\mu \in i} \int_0^\infty dt' h_i(t') \xi_\mu(t - t') \log[f_i(\vec{x} - \vec{x}_\mu)] \right\}.$$

**Optimized code for variance of motion which increases quadratically with time**

In this case Eq. (3) is replaced by (see SI)

(14)

$$\Delta^2 = 3 \cdot \left(\frac{v}{2J}\right)^{\frac{2}{3}}.$$

Consequently, the optimal number of neurons in module $n_i$ (Fig .5A) are given by (see SI)

(15)

$$n_i = \frac{\frac{1}{2}\left(\frac{2}{3}\right)^{m+1-i}}{1 - \left(\frac{2}{3}\right)^m} N,$$

the ratio $\lambda_i / \lambda_{i+1}$ (Fig .5C) is given by

(16)

$$\frac{\lambda_i}{\lambda_{i+1}} = \left[ \frac{\lambda_1 \beta^3 \alpha N}{3^{\frac{3}{2}} v \left(1 - \left(\frac{2}{3}\right)^m\right)} \right]^{\frac{1}{2}\left(\frac{2}{3}\right)^i} \left(\frac{3}{2}\right)^{\left[1 - \frac{1}{2}(m+3)\left(\frac{2}{3}\right)^i\right]},$$

and the optimal time constant for readout $\tau$ is given by

(17)

$$\tau = \left(\frac{1}{2Jv^2}\right)^{\frac{1}{3}}.$$

**SUPPLEMENTAL INFORMATION**

The supplemental Information is available from the authors on request.

**AUTHOR CONTRIBUTIONS**

**ACKNOWLEDGMENTS**

We thank Tor Stensola and Edvard Moser for permission to reproduce a figure from their paper, and Michael Hasselmo and Motoharu Yoshida for helpful correspondence. This research was supported by the Israel Science Foundation grant No. 1733/13 and (in part) by grant No. 1978/13. We acknowledge support from the Gatsby Charitable Foundation.

An earlier version of this work appeared in abstract form (Weiss et al., 2015).

**FIGURE LEGENDS**

**Figure 1. Modular organization and dynamic decoding. (A)** Experimental evidence for the modular organisation of grid cells [reproduced with permission of the authors from (Stensola et al., 2012)]: grid spacing in a single rat, where each dot corresponds to a single cell. Right, kernel smoothed density (KSD) estimate of the distribution. Red text, spacing in cm for the estimated peaks. Note the dramatic decline in the number of cells with larger grid spacings. See also additional Figs. in (Stensola et al., 2012). **(B)** Schematic illustration of the posterior distribution over position, inferred from spikes generated by all cells in a single module. The posterior has the same periodicity $\lambda$ as the single neuron tuning curves, the local variance of the peaks is denoted by $\Delta^2$. **(C)** Schematic illustration of a decoder for a dynamic variable, which follows the statistics of a simple random walk (shown for simplicity in one dimension). Black: a random walk trajectory $\vec{x}(t)$. Red lines represent spikes emitted by a population of neurons with different tuning curves, where the red y-axis represents the neuron index. The decoder estimates the animal's position at time $t_0$, based on all the spikes that occurred up to that time. **(D-E)** MSE's of an optimal decoder, estimating position based on spikes from a single module, as a function of the number of neurons, for a static two-dimensional variable **(D),** and a dynamic random variable, following the statistics of a simple random walk in two dimensions **(E)**. Blue dots: measurements of the MSE from simulations of an optimal decoder, responding to spikes generated by neural populations of varying size. Each dot represents an average over 300 realizations, where in **(D)** the averaging is over a single time interval from each simulation lasting 100 ms, and in **(E)** we average the MSE in each simulation also over time (realizations lasting at least ~200 ms) . Error bars: 1.96 standard deviations of the MSEs obtained from each simulation, divided by square root of the number of simulations. The receptive fields of the cells are Gaussians with $r_{max} = 10$Hz, $\lambda = 2.82$ m, $\Delta T = 100$ms for the static case **(D)**, and $D = 0.0125 \frac{\text{m}^2}{\text{s}}$, for the dynamic case **(E)**. Red lines: theoretical predictions from Eq. (1)-(2) **(D)** and Eq. (2)-(3) **(E)**.

**Figure 2. Optimized code: analytical results**. **(A)** Number of neurons in a module as a function of the module index (10 modules, ordered by grid spacing starting from the largest spacing). The total number of neurons is $N = 10^4$ (blue) and $N = 10^5$ (red). **(B)** Grid spacings in the optimized code. **(C)** Ratios between grid spacings in successive

modules. The ratio approaches $\sqrt{2}$ in the smallest modules. In all three panels, $\lambda_1 = 5\text{m}$, $D = 0.05\frac{\text{m}^2}{\text{s}}$, $\beta = 0.1$, and the receptive fields of the cells are Gaussians with $r_{max} = 10\text{Hz}$.

**Figure 3. Optimal Bayesian decoder**. The posterior probability distribution obtained using Eq. (10)- (Experimental Procedures) in simulations, shifted by the true position of the animal, and averaged over 1350 time points. Three different allocations of neurons to modules are shown: **(A)** optimal allocation as in Fig. 2A, **(B)** equal number of neurons in each module, and **(C)** reversed allocation. The MSE and margins of error noted on the left bottom of each panel were computed as in Fig. 1D-E, based on 100 simulations each lasting $\sim 1.4\text{s}$. All the parameters are the same as in Fig. 2, with $N = 10^4$.

**Figure 4. Simplified estimator**. **(A)** An illustration of the temporal exponential kernel, and the characteristic time $\tau$. A spike at time $t'$ contributes to the estimate of position at time $t$ with weight $e^{-\frac{t-t'}{\tau}}$. **(B)** Integration time constant $\tau$ as a function of the module index, as obtained from Eq. (7), substituting the values of $\lambda_i$ and $n_i$ from Fig. 2 ($N = 10^4$). **(C)** Schematic illustration of a model for readout (e.g. by place cell in the hippocampus). Each place cell approximate the log likelihood to be at a particular position given the spikes of multiple grid cells, as a linear summation of the spikes with integration times that vary depending on the grid spacing. **(D)** Performance of the simplified estimator, measured in the same way as in Fig. 3. All the parameters are the same as in Figs. 2 and 3, with $N = 10^4$.

**Figure 5. Optimization for other statistics of motion.** Optimized parameters for encoding and decoding by grid cells, as in Fig. 2 & Fig. 4B, but assuming that the variance of motion increases quadratically with time (see SI for details), and that readout is performed by an the simplified estimator. **(A)** Number of neurons in a module as a function of the module index. **(B)** The grid spacing. **(C)** Ratios between grid spacings in subsequent modules. This ratio approaches 1.5 in the smallest modules. **(D)** Integration time $\tau$ as a function of the module index, Eq. (17). All the parameters are the same as in Fig.2, with $N = 10^4$ and velocity $v = 1\text{m/s}$.

**Figure 1** (Weiss-Mosheiff et al.)

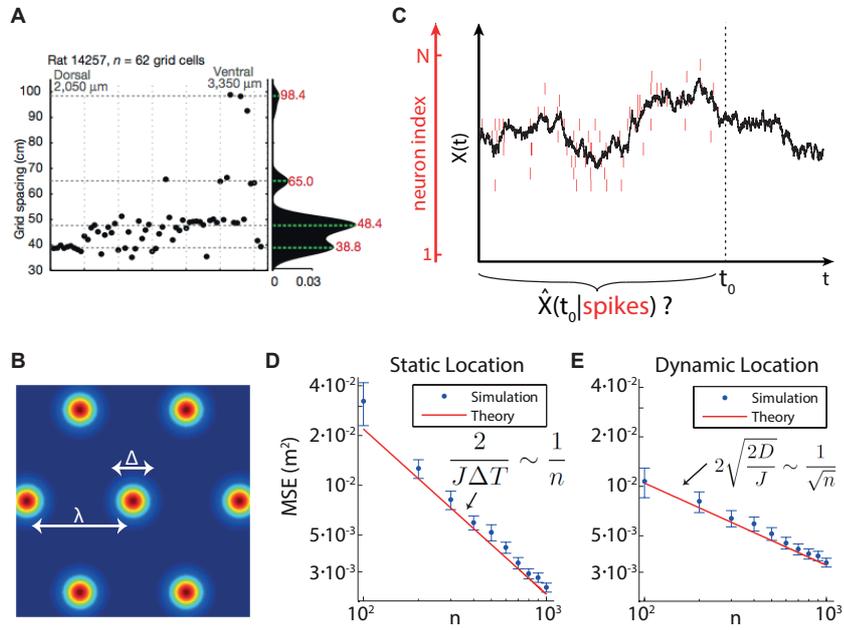

**A** Rat 14257, *n* = 62 grid cells

**B**

**C**

**D** Static Location

$$\frac{2}{J\Delta T} \sim \frac{1}{n}$$

**E** Dynamic Location

$$2\sqrt{\frac{2D}{J}} \sim \frac{1}{\sqrt{n}}$$

**Figure 2** (Weiss-Mosheiff et al.)

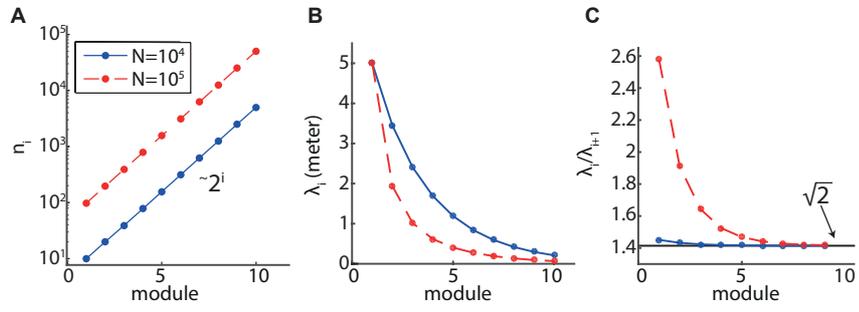

**Figure 3** (Weiss-Mosheiff et al.)

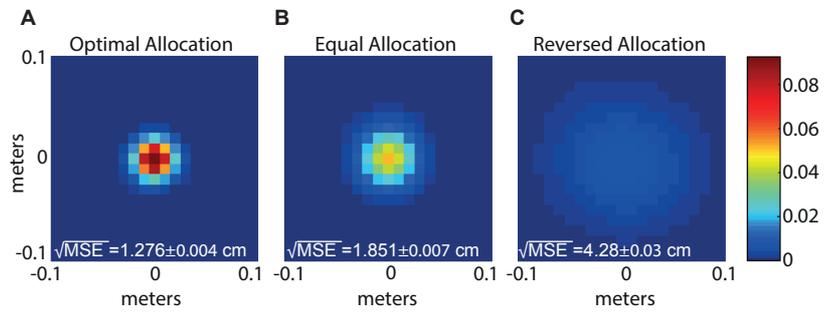

**Figure 4** (Weiss-Mosheiff et al.)

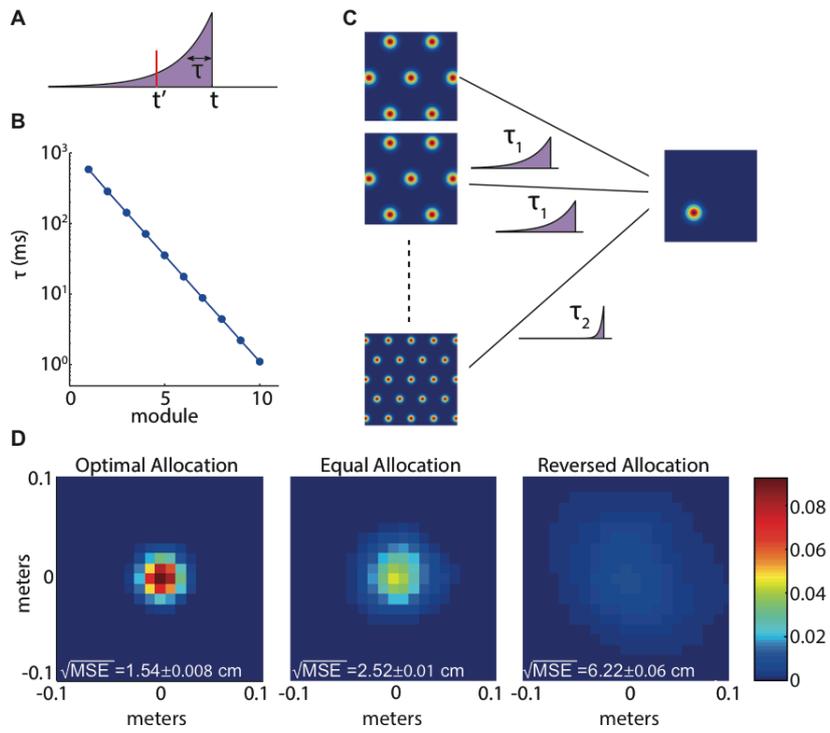

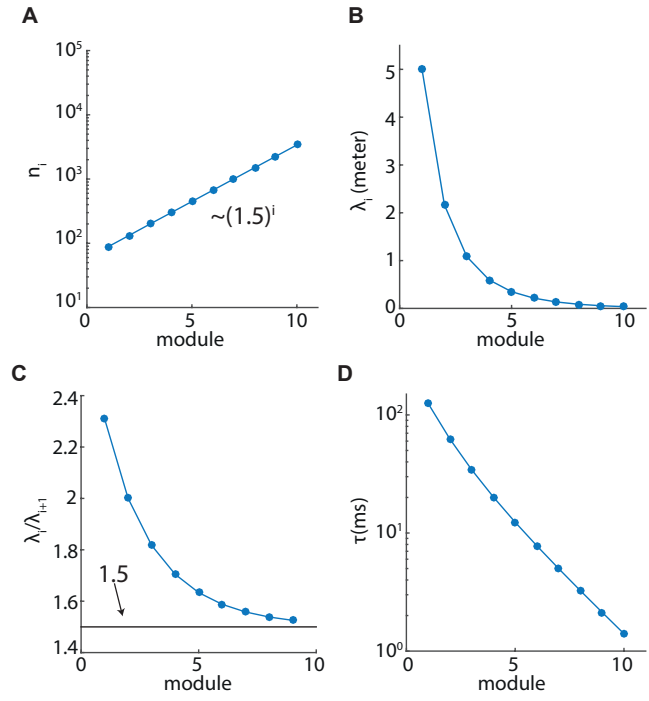